%Paper: hep-th/9512121
%From: Eric Zaslow <zaslow@abel.math.harvard.edu>
%Date: Sat, 16 Dec 1995 23:22:23 -0500

% This paper uses harvmac
\input harvmac.tex

\lref\rZtom{E. Zaslow, ``Topological Orbifold Models and Quantum
Cohomology Rings," Commun. Math. Phys. {\bf 156} (1993) 301-331.}
\lref\rDHVW{L. Dixon, J. Harvey, C. Vafa, and E. Witten,
``Strings on Orbifolds," Nucl. Phys. {\bf B261} (1985) 678-686,
and ``Strings on Orbifolds (II),"
Nucl. Phys. {\bf B274} (1986) 285-314.}
\lref\rVWodt{C. Vafa and E. Witten, ``On Orbifolds with Discrete Torsion,"
hep-th/9409188.}
\lref\rFHSVsqms{S. Ferrara, J. Harvey, A. Strominger, and C. Vafa,
``Second-Quantized Mirror Symmetry," hep-th/9505162.}
\lref\rSVdp{A. Sen and C. Vafa, ``Dual Pairs of Type II String
Compactification,"
hep-th/9508064.}
\lref\rVe{I. Vainsencher, ``Enumeration of $n$-Fold Tangent Hyperplances to a
Surface," J. Alg. Geom {\bf 4} (1995) 503-526.}
\lref\rBSVII{M. Bershadsky, V. Sadov, and C. Vafa, ``D-Branes and Topological
Field Theories," hep-th/9511222.}
\lref\rVst{C. Vafa, ``A Stringy Test of the Fate of the Conifold,"
Nucl. Phys. {\bf B447} (1995) 252-260.}
\lref\rBSVI{M. Bershadsky, C. Vafa, and V. Sadov, ``D-Strings on
D-Manifolds," hep-th/9510225 (revised version).}
\lref\rOVtdbh{H. Ooguri and C. Vafa, ``Two-Dimensional Black Hole and
Singularities of CY Manifolds," hep-th/9511164.}
\lref\rPdb{J. Polchinski, ``Dirichlet-Branes and Ramond-Ramond Charges," hep-th
9510017.}
\lref\rEOYscfa{T. Eguchi, H. Ooguri, and S.-K. Yang, ``Superconformal Algebras
and String Compactification on Manifolds with $SU(n)$ Holonomy," Nucl.
Phys. {\bf B}315 (1989) 193-221.}
\lref\rSmsbs{A. Sen, ``A Note on Marginally Stable Bound States in Type II
String Theory," hep-th/9510229.}
\lref\rSud{A. Sen, ``U-Duality and Intersecting D-Branes," hep-th/9511026.}
\lref\rHMabs{J. Harvey and G. Moore, ``Algebras, BPS States, and Strings,"
hep-th/9510182.}
\lref\rWstd{E. Witten, ``String Theory Dynamics in Various Dimensions," hep-th
9503124.}
\lref\rDELhds{R. Donagi, L. Ein, and R. Lazarsfeld, ``A Non-linear Deformation
of the Hitchin Dynamical System," alg-geom 9504017.}
\lref\rVg{C. Vafa, ``Gas of D-Branes and Hagedorn Density of BPS States,"
hep-th/9511088.}
\lref\rDHnst{A. Dabholkar and J. Harvey, ``Nonrenormalization of the
Superstring Tension," Phys. Rev. Lett. {\bf 63} (1989) 478-481.}
\lref\rNnhs{K. S. Narain, ``New Heterotic String Theories in Uncompactified
Dimensions $< 10,"$ Phys. Lett. {\bf B169} (1986) 41-46.}
\lref\rNSWn{K. S. Narain, M. H. Sarmadi, and E. Witten, ``A Note on
Toroidal Compactification of Heterotic String Theory," Nucl Phys.
{\bf B279} (1987) 369-379.}
\lref\rWbspb{E. Witten, ``Bound States of p-Branes," hep-th/9510135.}
\lref\rBBS{K. Becker, M. Becker, and A. Strominger, ``Fivebranes, Membranes,
and Non-Perturbative String Theory," hep-th/9507158.}
\lref\rWbs{E. Witten, ``Bound States of Strings and p-Branes," hep-th/9510135.}
\lref\rBJSVtop{M. Bershadsky, A. Johansen, V. Sadov, and C. Vafa,
``Topological Reduction of 4D SSYM to 2D $\sigma-$Models," Nucl.
Phys. {\bf B448} (1995) 166.}
\lref\rCcount{X. Chen, work in progress.}
\lref\rVWsct{C. Vafa and E. Witten, ``A Strong Coupling Test of $S-$Duality,"
Nucl. Phys. {\bf B431} (1994) 3.}
\lref\rVinst{C. Vafa, ``Instantons on D-Branes," hep-th/9512078.}
\lref\rLkIII{J. Wolfson, ``Minimal Surfaces in K\"ahler Surfaces
and Ricci Curvature," J. Diff. Geom. {\bf 29} (1989) 281-294.}
\lref\rCthesis{L. Caporaso, ``A Compactification of the Universal
Picard Variety over the Moduli Space of Stable Curves," JAMS {\bf 7}
(1994) 589-660.}
\lref\rHsl{R. Harvey and
H. B. Lawson, Jr., ``Calibrated Geometries,"
Acta Math. {\bf 148} (1982) 47-157;
R. Harvey, {\sl Spinors and Calibrations,} Academic
Press, Boston, 1990.}
\lref\rDual{C. M. Hull and P. K. Townsend, ``Unity of
Superstring Dualities," Nucl. Phys. {\bf B438} (1995) 109-137;
A. Sen, ``Strong-Weak Coupling Dulaity in Four-Dimensional
String Theory," Int. J. Mod. Phys. {\bf A9} (1994) 3707-3750;
J. Schwarz, ``Does String Theory Have a Duality Symmetry
Relating Weak and Strong Coupling?" Proceedings, Strings
'93, Berkeley, 1993.}
\lref\rPDLnc{J. Dai, R. G. Leigh, and J. Polchinski,
Mod. Phys. Lett. {\bf A4,} (1989) 2073.}
\lref\rKVer{S. Kachru and C. Vafa, ``Exact Results for N=2 Compactifications
of Heterotic Strings," Nucl. Phys. {\bf B450} (1995) 69-89.}
\lref\rGHalg{P. Griffiths and J. Harris, {\sl Principles of Algebraic
Geometry,} Wiley-Interscience, New York, 1978.}

%% Some definitions

\def\sub{\scriptscriptstyle}

\def\pg{{\bf P}^g}
\def\moduli{{\cal M}_g^H}
\def\zbar{\overline{z}}
\def\vac{\vert 0\rangle}

\Title{}{\vbox{\centerline{BPS States, String Duality, and Nodal Curves
on K3}}}
\centerline{Shing-Tung Yau and Eric Zaslow}
\vskip 0.25in
{\sl Mathematics Department, Harvard University,
Cambridge, MA 02138, USA}

\vskip 0.5 in

We describe the counting of BPS states of Type II on K3
by relating the supersymmetric cycles of genus $g$ to the
number of rational curves with $g$ double points on K3.
The generating function for the number of
such curves is the
left-moving partition function of the bosonic string

\Date{12/95}

\newsec{Introduction}

The revolution afoot in string theory bespeaks the promise of understanding
non-perturbative strings.  Solitonic states of the low-energy physics \rBBS\
have been interpreted as simple conformal field theories of open strings \rPdb.
This observation has had dramatic consequences, most strikingly in
helping to provide evidence for string duality, wherein the non-perturbative
physics of one theory is equated with the perturbative regime of another
\rWstd\rDual.
The hallmark for demonstrating such equalities has been a counting
of BPS states, which are the signatures of a supersymmetric theory.
D-branes in Type II string theory, equivalent to the
so-called supersymmetric cycles,
are precisely such states.  The minimum energy states in the presence of
a D-brane can be attained by solving for the vacua of the effective
low-energy quantum field theory, which is a counting of the number of
such BPS states.

This problem, for Type II
compactifincation on K3, has been detailed in
a recent paper \rBSVII, in which the authors equate this problem to a
supersymmetric
quantum field theory with target space equal to a certain moduli space.
(For the problem on $T^4,$ see also \rSud\rSmsbs\rVg.)
The moduli space is that of a supersymmetric cycle,
or in mathematical terms ``special Lagrangian manifold,'' with a line
bundle on K3 -- which they argue
to be cohomologically equivalent to symmetric products of K3 itself.
In this paper, we show that this moduli space problem is
equivalent to the mathematical question of counting the number of
rational curves in K3 with $n$ double points (where pairs of points
are mapped to the same point).  We count these by considering the
hyperplanes in a projective space ${\bf P}^n$ which intersect an
embedded K3 at a Riemann surface of genus $n.$
Quite strikingly, the string duality equating Type II on K3 to the
heterotic string on $T^4$ provides a very simple explanation of this
counting, yielding a generating function for all such numbers for
rational curves of arbitrary topological genus -- it is none other than
the partition function of the bosonic string!  The first six coefficients
agree with the current results known by mathematicians \rVe; the remaining
coefficients can be taken as explicit conjectures for further
verification.  This counting is currently being pursued \rCcount.

In section two we review the relevant aspects of string theory which
lead to the conclusion that special Lagrangian submanifolds of a
compactifying space correspond to BPS solutions of the effective
supergravity theory, or D-branes.  We review the work of
Ref. $\!\!\rBSVII,$ which gives the physical counting of states.  The
heterotic duality gives an equivalent counting.  In section three
we describe why this counting is equivalent to the mathematical
question described above,
and report the results of Ref. $\!\!\rVe.$  In section four we
conclude with some conjectures and prospects for further study.

\newsec{String Theory, BPS States, and D-Branes on K3}

The low energy description of Type II string theory is a supergravity theory.
Recently, Becker et al. have looked at solitonic solutions of this theory
which preserve half the supersymmetries \rBBS.

A supersymmetric three-cycle is defined as a three-cycle solution of eleven
dimensional supergravity,
compactified on a Calbi-Yau,
which preserves one-half of the supersymmetries
(which are generically totally broken).
This was shown to be a good definition for the
Type II theory \rBBS, whose strong coupling behavior is believed to be
the eleven dimensional supergravity \rWstd.  The condition leads to the
definition that $i: C \hookrightarrow X$ is an embedding of a
supersymmetric cycle ($\hbox{dim}_{\bf R}C = 3;$ $\hbox{dim}_{\bf R}X = 6$)
if
$$i^*\Omega \sim V,\qquad\qquad i^*\omega = 0,$$
where $\Omega$ is the holomorphic three-form, $\omega$ the K\"ahler class,
$V$ the volume form induced from the embedding, and the proportionality is
described by a multiple which is constant -- not a function -- on $C.$  (This
constant can be set to one at a point in moduli space
by a rescaling of $\Omega.$)  The second condition is that of a Lagrangian
suface, and the two together describe ``special Lagrangian"
submanifolds \rHsl.  Unfortunately,
although they are closely related to
the much-studied minimal submanifolds, very little is known about
such submanifolds in non-trivial spaces.

In the present case, we study cycles which have two dimensions in a K3,
and the rest flat (${\bf R}\times S^1$).
We consider the worldvolume of
a spatial three-brane, which will have a four-dimensional effective theory of
the low-energy physics.
The two-manifold in K3 will be a special
Lagrangian submanifold, and such were proven \rLkIII\ to be
equivalent to curves which are holomorphic with respect to one of the
complex structures on K3 (see also the argument in
\rBSVII).\footnote{$^{*}$}{Recall that the hyperkahler structure
gives us an $S^2$ of complex structures from the three
independent, quaternionic
$J_i$: $J = aJ_1 + bJ_2 + cJ_3$ is a complex structure
for $a^2 + b^2 + c^2 = 1.$}

The thrust of Ref. $\!\!$\rBSVII\ was to study the effective theory in the
presence of a D-brane.  As D-branes are BPS states, they represent
minimum-energy states in a given topological sector (those which
saturate the Bogomol'nyi bound).  It has been argued \rPDLnc\rWbs\ that
the low energy field theory limit of $n$ coincident
D-brane strings was a $U(n)$ gauge theory reduced to the full
dimension of the $D-$brane (including time).
This comes about from the $n$ $U(1)$'s of the open string spectra
plus the now massless $n(n-1)$ strings stretching between different
$D-$branes.  As discussed above,
we enlarge the space by a circle (converting Type IIA to IIB) so
that we may think of the three-brane as a two-surface, $\Sigma,$
crossed with $S^1\times{\bf R},$ a four-dimensional space.
In fact, it is a twisted supersymmetric field theory,
precisely of the type considered in Ref. $\!\!\rBJSVtop,$ so its ground
states can be obtained by reduction from four to two
dimensions, yielding a two-dimensional supersymmetric sigma model
on $S^1\times {\bf R}.$
What is the target space of this sigma model?

As shown in \rBSVII\ and \rBJSVtop, the target space is the space of solutions
to the equations
which must be solved for the dimensional reduction to make sense -- the
Hitchin equations:
$$\matrix{F_{z\zbar} = [\Phi_z,\Phi_{\zbar}]\cr
&\cr
D_z\Phi_{\zbar} = 0 = D_{\zbar}\Phi_z.}$$
Here $F$ is the gauge field strength and $\Phi$ is an adjoint-valued
one-form -- identified with the normal vector
in the dimensional reduction of the gauge field, which is possible
due to the codimension one embedding in a space with
trivial canonical bundle, the K3.
For our purposes, we will only consider one D-brane ($n = 1$),
so that $F = 0$ and we have the space of flat connections and
a choice of harmonic one-form.  The one-form describes motions in
the normal direction, so is equivalently giving a local parametrization
of the space of holomorphic surfaces of genus $g.$  The ``compactification"
of the space of solutions to the Hitchin equations then yields a moduli
space, $\moduli,$ describing a choice of holomorphic Riemann surface
in K3 and a flat $U(1)$ bundle (a point in the Jacobian).

As the authors of Ref. $\!\!$\rBSVII\ argue, this space is
at least birational to a symmetric product of K3 itself.
Since the ground states of a supersymmetric sigma model are
known to be equivalent to the set of all cohomogy classes,
it remains to compute the total cohomology space\footnote{$^{**}$}{Not
to be confused with ``total cohomology," a term used for
double complexes.} of this symmetric product.  Such a counting
is made simple by orbifold techniques \rDHVW\rZtom\
(believed exact for hyperkahler manifolds), and was computed
in Ref. $\!\!\rVWsct.$  Let us briefly
review this computation.

The cohomology of the symmetric product is
given by the oscillator level $g$ for bosonic and fermionic
oscillators $\alpha_{-k}^i$ acting on a Fock vacuum $\vac,$ where
$k = 1...g$ and $i = 1...\hbox{dim}H^*(M),$ one for each
cohomology element -- with the bose/fermi statistics
depending on whether it is an odd or even cohomology class.
The reason this is so is that the orbifold
cohomology contains the $S_g-$invariant cohomology of $M^{\otimes n},$
i.e. the properly symmetrized polynomials in $\hbox{dim}H^*(M)$ variables,
equivalent to states of the form
$$\alpha_{-1}^{i_1}...\alpha_{-1}^{i_g}\vac.$$
Here all the oscillators are bosonic for K3 and so the corresponding
polynomials are totally symmetric.
The twisted sectors, one for each conjugacy class of $S_g$ account for
the rest of the oscillators.  The conformal weights of the
oscillators in the twisted sectors
are determined by the order of the cyclic pieces of
the permutations
(all are factorable into products of cyclic permutations)
in a conjugacy class -- see \rVWsct\ for details.
By this reasoning, we can compute all the dimensions of the cohomology
rings the $\hbox{Sym}^g(M)$ at once, and organize them in a generating
function.
The result for K3, which has no cohomlogy classes of odd degree
(and hence no ferminic oscillators), is
that
$$\sum_{g = 1}^{\infty}q^g\,\hbox{dim}H^*(\hbox{Sym}^g(\hbox{K3})) =
\prod_{g = 1}^{\infty}{1\over (1-q^g)^{24}} = q\eta(q)^{-24}.$$

In fact, BPS states have a much simpler description \rVg\rVinst,
if one appeals
to the string duality relating the Type II string on K3 to the heterotic
string on $T^4,$ defined most conveniently via the Narain lattice
\rNnhs\rNSWn.
In the heterotic theory, the BPS states have a simple description \rSud:
the right-moving oscillators must be in their ground states.  The
right-moving momentum, $p_R,$ yields the mass;
the left oscillator number gives the topological
type $(p^2)$ of the BPS state.  In the Type II language we have
that the self-intersection
number, equivalently the topological genus, of the curve is
equal to the left oscillator level. But since
left-movers are just free bosons, the number $d_g$
of states at oscillator level
$g$ is just the coefficient of $q^g$ in the partition function of
the bosonic string (ignoring the $q^{-c/24}$ piece).  This is the
same as the result stated above! (See \rVinst, too.)  For example, at
ocillator level 3 we have states of the form
$$\alpha_{-1}^i\alpha_{-1}^j\alpha_{-1}^k\vac,\qquad
\alpha_{-2}^i\alpha_{-1}^j\vac,\qquad \alpha_{-3}^i\vac.$$
The number of states of the first type is the number of symmetric
polynomials of degree three in twenty-four variables (the transverse
modes of the string), or $\pmatrix{24 + 3 - 1\cr 3} = 2600.$  Thus
$d_3 = 2600 + (24)^2 + 24 = 3200.$
Higher levels can be obtained similarly, or by exanding the
infinite product in a power series.  One finds:
$$\eqalign{q(\eta (q))^{-24} &=
\prod_{n=1}^{\infty}{1\over (1-q^n)^{24}} \cr
&= \sum_{g = 0}^{\infty}d_g q^g \cr
&= 1 + 24q + 324q^2 + 3200q^3 + 25650q^4
+ 176256q^5 + 1073720q^6 + ...}$$
and
\eqn\one{d_g = \chi(\moduli).}
These numbers agree precisely with those known by mathematicians for
the numbers $N_g$ of rational $g-$nodal
curves in K3, a computation
to which we now turn.  The numbers $N_g$ for
$g > 6$ are not known mathematically.

\newsec{Nodal Curves in Linear Systems}

In order to compare with computations done in mathematics, we
relate the physical question to an equivalent mathematical one:
how many ${\bf P}^1$'s with $g$ double points sit inside K3?
Or,
how many hyperplanes, among a linear system
in ${\bf P}^g,$ intersect an embedded K3 in $g$ double points?
In order to understand the equivalence, it is helpful to
elaborate on what is being asked through a simple example.

Consider a K3 embedded in ${\bf P}^3,$ expressed as the zeros of
a degree four homogeneous polynomial (here of Fermat type)
$$X^4 + Y^4 + Z^4 + W^4 = 0,$$
where $(X,Y,Z,W)$ are homogeneous coordinates
(the results will be independent of the complex structure of K3,
so we choose a simple embedding).
A linear system
is a set of hyperplanes, or zeros of a linear polynomial.
We can parametrize a hyperplane by the complex coefficients
$\alpha = (a,b,c,d)$ of the linear polynomial:
$$H_{\alpha} = \{ aX + bY + cZ + dW = 0 \} \subset {\bf P}^3.$$
Clearly the space of inequivalent $\alpha$ defines a ${\bf P}^4,$
which we call ${\bf P}^{4*}$.
Now $H_\alpha$ intersects the K3 along the curve (say, when $d \neq 0$)
$$P_\alpha = d^4X^4 + d^4Y^4 + d^4Z^4 + (-aX -bY - cZ)^4 = 0.$$
The adjunction formula (physicist's version)
tells us that this curve has (topological)
genus $g = 3,$ as expected.  However, not all such curves are non-singular.
Whether the curve is singular depends on whether the equations
$$P_\alpha = dP_\alpha = 0$$
(here $d$ is exterior differentiation)
have simultaneous solutions other than
$X = Y = Z = 0.$  This clearly depends on the modulus $\alpha.$
In fact, the simultaneous equations can be written as the
discriminant locus of a larger equation, whose solutions correspond
to the points $\alpha$ describing singular submanifolds.
The set of all $\alpha \in {\bf P}^{4*}$
such that this intersection is a rational curve with $3$ (in general
$g$) nodes (i.e. double points) defines a subset of ${\bf P}^{4*}.$
For the case considered, this subset is a finite point set with
a number of points $N_g$ to be determined.

Why is this number equal to the dimension of the cohomology
space of the
moduli space, $\moduli,$ of supersymmetric cycles of genus $g$
with line bundles
(and given homology type)?
First, since all the cohomology elements occur with
even dimension (as we saw in the last section),
we are simply computing the Euler
characteristic of $\moduli.$  This was, perhaps, a
fortunate coincidence; for generalizations of this
problem one would have to determine whether it makes sense
to count the BPS states with or without signs to relate to mathematical
computations.  For example, $T^4$ has $\chi(T^4)=0,$ but indeed
has BPS states; so either the mathematics is more naturally
associated to the full cohomology ring, or the would-be
invariants are trivial.
Happily neglecting such subtleties, we prove equivalence to
the Euler characteristic for the case at hand.

Consider a K3 embedded in a $\pg$ such that $L,$ the pull-back of
the hyperplane line bundle has $h^0 = g + 1,$ i.e. all the
global sections -- the homogeneous coordinate functions $X^i$ --
are preserved.  It is clear here that a choice of global sections
of $L$ define the embedding, up to projective transformations.
This describes the canonical embedding into ${\bf P}(H^0(\hbox{K3},L)).$
The zeros of the global sections $\sum a_iX^i$
define divisors, equivalent
to hyperplanes intersecting the K3.
Let $\Sigma$ be such a divisor, so that $L = [\Sigma].$
The ${\bf P}^{g*}$ of such divisors is called the linear system.
Describing the hypersurfaces locally as the zeros of
a defining polynomial, one finds for the normal
bundle, $N_\Sigma = [\Sigma]\vert_\Sigma.$  Then the exact sequence
$$0\longrightarrow T\Sigma\longrightarrow T\hbox{K3}\vert_\Sigma
\longrightarrow N_\Sigma\longrightarrow 0$$
tells us, upon dualizing (which reverses arrows)
and taking the exterior product,
\eqn\zero{K_{\Sigma} = (K_{\rm{K3}} + [\Sigma])\vert_\Sigma.}
Here $K$ denotes the canonical bundle (the highest exterior
power of the holomorphic cotangent bundle) and additive
notation is used for tensor products.
The Riemann-Roch theorem for surfaces \rGHalg\
relates the Euler characteristic
of a line bundle to properties of the variety:
$$\chi([\Sigma]) = \chi({\cal O}_{\rm K3}) + ([\Sigma]\cdot
[\Sigma]-K_{\rm K3}\cdot[\Sigma])/2.$$
Here the dot product is integration of the first Chern class of
the line bundle.  Now $\chi({\cal O}_{\rm K3}) =
h^{0,0}-h^{1,0}+h^{2,0} = 2,$ and $K_{\rm K3}\cdot[\Sigma] = 0$
since the canonical bundle is trivial.
On the left hand side,
we have $H^p([\Sigma])=0,$ $p>0,$ by the
Kodaira vanishing theorem, since $K_{\rm K3}$ is trivial and
$[\Sigma]$ has global sections. Thus $\chi([\Sigma]) = g+1.$
Switching the sign of $K_{\rm K3}\cdot[\Sigma]$ on the right
hand side, and using \zero, we see
$$K_\Sigma\cdot [\Sigma] = 2g - 2,$$
i.e. $\Sigma$ has topological genus equal to $g.$
As the generic K3 has
a one-dimensional Picard group of line bundles
(obtained from the embedding), we will count all the
holomorphic curves this way; if there were more, there would be
another divisor class.

The key point now is that the space $\moduli,$ defined as the
(compactification of the) space of solutions to the Hitchin
equations describing the space of holomorphic
curves with choice of flat bundle, fibers over the
space of holomorphic curves on K3, with fiber equal
to the Jacobian of the curve.  The map is simple -- just
forget about the bundle information:
$$\matrix{\hbox{Jac}(\Sigma_\alpha)&\longrightarrow&\moduli\cr
{}&{}&\downarrow\cr
{}&{}&{\bf P}^{g*},}$$
where $\Sigma_\alpha$ is the curve described by $\alpha \in {\bf P}^{g*}.$
A non-singular curve will have non-singular Jacobian, isomorphic
to a torus $T^{2g}.$  A singular curve, say one
described by $\alpha_i \in {\bf P}^{g*},$
will have a fiber $F_i$ equal to a ``singular," or
more precisely, compactified,  Jacobian.
Therefore to compute the Euler characteristic, we can use
the simple fact that the Euler characteristic is additive.
Let $S = \bigcup S_i$ be the union of components $S_i \subset
{\bf P}^{g*}$ describing singular curves.
Then we have
\eqn\two{\eqalign{\chi(\moduli)& =
\chi({\bf P}^{g*}\setminus S)\cdot\chi(T^{2g})
+ \sum_i\chi(S_i)\cdot\chi(F_i)\cr
&= 0 + \sum_i \chi(S_i)\cdot\chi(F_i)}.}
Generically, curves with
$g-k$ holes and $k$ nodes
exist on a subvariety of codimension $k$ in
${\bf P}^{g*}.$  The ${\bf P}^1$'s with nodes exist on a finite
point set.

Let us consider what types of singular curves will result.
The fibers over the singular curves result from a Mumford compactification
of the Hitchin space (which is the total family of Jacobians of the
holomorphic curves) and correspond to compactifications of the
generalized Jacobian of singular spaces.  These sorts of Jacobians
were considered in \rCthesis.
We shall work only with nodal singularities or double points (which
look locally like $y^2 = x^2$).  The problem of compactifying
the generalized Jacobian over higher singularities is
poorly understood at this time; we hope that a better mathematical
understanding will corroborate our conjecture that they have
zero Euler characteristic, which we compute here for the more
common nodal singularities.
We now show that the Euler characteristic is zero for
all compactified Jacobians over Riemann surfaces with nodes,
except for those describing
curves of genus zero with $g$ double points.  These Jacobians
all have Euler characteristic equal to one.\footnote{$^{***}$}{We thank
L. Caporaso for explaining the compactified Jacobians and their
stratification.}

Consider first a ${\bf P}^1$ with one node; call it $X$.  Its
normalization is a ${\bf P}^1$ with no nodes.  One constructs a flat
bundle over $X$ by considering a flat bundle over ${\bf P}^1$ and
then identifying the fibers over two points sitting over the double
point.  The identification of the fibers gives a ${\bf C}^*$ ambiguity,
so that the generalized Jacobian, $J,$ sits inside a sequence
$$0 \longrightarrow {\bf C}^* \longrightarrow J \longrightarrow
\{\hbox{pt}\} \longrightarrow 0,$$
where $\{\hbox{pt}\}$ is just a point,
representing the Jacobian of ${\bf P}^1.$
The compactification $\overline{J}$ of $J$ is a one-point (one-stratum)
compactification corresponding to the
(unique up to isomorphism) line bundle over
a blowup, up to isomorphism,
obtained by an exceptional divisor intersecting the
normalization at the two points to be identified.  This line bundle
must be degree one on the exceptional divisor, and degree $-1$ on
the normalization -- the total degree must be zero, as we are describing
a Jacobian.  There is one such line bundle, since the normalization
${\bf P}^1$ has trivial Jacobian.
The result is a one-point compactification of ${\bf C}^*,$
which, coincidentally, can be thought of again as ${\bf P}^1$ with a
node.  What is important to us is the stratification.
Let us calculate $\chi(\overline{J})$ by summing up the
pieces.  First we have
an open set ${\bf C}^*\times \{\hbox{pt}\},$
representing a line bundle over a point
(in general, this point will be replaced by the Jacobian of the
normalization).  This has Euler characteristic zero.  Then we
have the single stratum which is equal to a point, with Euler
characteristic equal to one.  The total Euler characteristic
is one.

For $d>0$ nodes on a ${\bf P}^1$
the proof is similar, with ${\bf C}^*$ getting replaced by
$({\bf C}^*)^d,$ corresponding to
gluing choices of $d$ pairs of points on the
normalization of the curve.
A dense open subset of $\overline{J}$ is always
obtained as a $({\bf C}^*)^d$ bundle over the Jacobian of the
normalization.  This has Euler
characteristic zero.  The lowest stratum is a point corresponding to a
{\sl unique} line
bundle with prescribed degrees
over the blowup with exceptional divisors intersecting paired
points -- which has Euler characteristic one.
For higher genus Riemann surfaces with nodes,
there is again a stratification, and the proof is again by
induction.  Zero always results, for the strata are either
spaces of line bundles over Riemann surfaces -- in which case
the Jacobian is a non-trivial torus, yielding zero Euler
characteristic -- or are higher-genus surfaces with
fewer singularities, which will fall under the inductive hypothesis.
${\bf P}^1$ is special because it has a unique flat bundle.
For higher types of singularities, which can occur, we expect
a zero contribution.

Summarizing, we found for nodal singularities
$$\sum_i\chi(S_i)\cdot\chi(F_i) =
\sum_{\matrix{\sub{\rm{genus}\; p>0},\cr
{}^{\sub{g-p \;\rm{nodes} }} }}
{\chi(S_j)\cdot\chi(\overline{J}_{S_j})} +
\sum_{\matrix{\sub{\rm genus\; 0},\cr
{}^{\sub{g \;\rm nodes}} }}
\chi(\{\alpha_j\})\cdot\chi(\overline{J}_{\alpha_j}),$$
where in the first term $\chi(\overline{J}_{S_j})=0$
for all types of singular curves labeled by $S_j,$ and
$\chi(\{\alpha_j\})= \chi(\overline{J}_{\alpha_j}) = 1$
for the singular curves
in the second sum.  Thus, recalling \one\ and \two, we have
\eqn\three{\chi(\moduli) = N_g = d_g.}

\newsec{Conclusions and Prospects}

We hope this argument leads
to further mathematical confirmation of the observations in this paper.
In particular, we conjecture that $N_7 = 5930496, N_8 = 30178575,$
and so on.  These numbers, defined rigorously, can
be considered as tests of string duality or of the strength of
the arguments in deriving the effective theories of D-branes.  That
agreement has, to date, been achieved up to $g = 6$ is already quite
striking.  Note that for $g \leq 2$ although there is no embedding of the
K3 in projective space $\pg,$ it makes perfect sense to consider
the spaces of holomorphic curves of genus $g \leq 2.$
The $g = 1$ case, for example, corresponds to the map
$\hbox{K3} \rightarrow {\bf P}^1,$ viewing K3 as an
elliptic fibration over ${\bf P}^1$ with 24 singular fibers.
For $g \geq 7,$ we eagerly await mathematical computations.

Unfortunately, both the physical and mathematical pieces of the puzzle
are quite difficult.  What made this problem tractable was 1) the
identification of the moduli space around a supersymmetric cycle,
made possible essentially by K3 being a toric fibration; 2) the simple
dual string computation; 3) the relation, in two dimensions, between
supersymmetric cycles and holomorphic maps.  As K3 and $T^4$ are the
only viable four-dimensional compactifying spaces, generalizations
of the above lead us to higher-dimensional spaces.  For Calabi-Yau
compactifications, special Lagrangian manifolds are (real)
three-dimensional, and so we lose any connection to holomorphicity.  It
is unlikely, therefore,
that the computation has a simple algebro-geometric description.
Supersymmetric four-cycles (not special Lagrangian) in Calabi-Yau's
may have a holomorphic interpretation, but tests of string
duality and quantum mirror symmetry \rKVer\rFHSVsqms\ seem to
require a better mathematical understanding of special Lagrangian
embeddings.

\bigbreak\bigskip\bigskip\centerline{{\bf Acknowledgements}}\nobreak

We would like to thank L. Caporaso, X. Chen, J. Harris, B. Lian,
A.-K. Liu, S. Srock, and C. Vafa
for discussions.  This work is supported
by grant
DE-F602-88ER-25065.

\listrefs

\end